\documentclass[a4paper,fleqn,usenatbib]{mnras}

\usepackage{newtxtext,newtxmath}

\usepackage[T1]{fontenc}
\usepackage{ae,aecompl}

\usepackage{graphicx}	
\usepackage{amsmath}	
\usepackage{amssymb}	

\title[Catastrophic disruption threshold of pre-planetary matter]{The tensile strength of compressed dust samples and the catastrophic disruption threshold of pre-planetary matter}

\author[I. L. San Sebasti\'an et al.]{
I. L. San Sebasti\'an,$^{1,2}$\thanks{E-mail: irina@fcaglp.unlp.edu.ar}
A. Dolff,$^{3}$
J. Blum,$^{3}$
M. G. Parisi, $^{1,4}$
S. Kothe $^{3}$
\\
$^{1}$Facultad de Ciencias Astron\'omicas y Geof\'isicas, Universidad Nacional de La Plata, Paseo del bosque s/n, La Plata 1900, Argentina\\
$^{2}$Instituto de Astrof\'isica de La Plata, CCT La Plata-CONICET-UNLP, Paseo del Bosque s/n, 1900 La Plata, Buenos Aires,  Argentina\\
$^{3}$Institut für Geophysik und extraterrestrische Physik, Technische Universität Braunschweig, Mendelssohnstr. 3, 38106 Braunschweig, Germany\\
$^{4}$Instituto Argentino de Radioastronom\'{ \i}a, CCT La Plata- CONICET- CICPBA, CC N$^{o}$5 Villa Elisa, Buenos Aires, Argentina\\
}

\date{Accepted XXX. Received YYY; in original form ZZZ}

\pubyear{2019}

\begin{document}
\label{firstpage}
\pagerange{\pageref{firstpage}--\pageref{lastpage}}
\maketitle

\begin{abstract}

During the planetary formation process, mutual collisions among planetesimals take place, making an impact on their porosity evolution. The outcome of these collisions depends, among other parameters, on the tensile strength of the colliding objects. In the first stage of this work, we performed impact experiments into dust samples, assembled with material analogous to that of the primitive Solar System, to obtain highly compressed samples that represent the porosities measured in chondritic meteorites. In the second stage, we obtained the tensile strengths of the compressed dust samples by the Brazilian Disk Test. We found a correlation between the tensile strength and the volume filling factor of the compressed dust samples and obtained the corresponding critical fragmentation strength in mutual collisions and its dependence on the volume filling factor. Finally, we give prescriptions for the catastrophic disruption threshold as a function of the object size, for different values of the volume filling factor that can be utilized in collisional models.

\end{abstract}




\section{Introduction}
\label{sec1}

Planetesimals, as well as pebble-sized dust aggregates, collide with each other during the planetary formation process. Depending on the particle sizes and impact velocities, these collisions may result in different outcomes, such as growth, rebound, erosion, mass transfer, fragmentation, craterization, compaction, or grinding \citep{blum2018}. At present, whether primordial planetesimals are born small or big remains a matter of debate \citep{Morbidelli2009,Weidenschilling2011}. The difference between both cases resides mainly in the impact-velocity regime in which collisions among planetesimals take place
\citep{SanSebastian2019}. Assuming a Keplerian shear regime without significant gravitational stirring, collisions among planetesimals occur at low impact speeds and the outcome of an impact between two small planetesimals leads to growth \citep{Weidenschilling2011}. In the past, it was usually assumed that initially 1-10~km-sized planetesimals were formed through coagulation processes in the protoplanetary disk and continued to agglomerate via pairwise mergers. However, it was shown that the dynamical evolution of the protoplanetary disk results in the dispersion-dominated regime, in which the relative velocity among small planetesimals increases due to the gravitational perturbations produced by planetary embryos, leading to disruptive collisions among them \citep{Morbidelli2009,Parisi2013,Sanse2016}.

These processes influence the porosity evolution of the planetesimals. The porosity, $\psi$, is related to the volume fraction occupied by the material, $\phi$, by $\phi = 1 - \psi$. Asteroids and comets are leftover planetesimals that provide us with evidence of their collisional evolution through their physical and dynamical parameters. Chondritic meteorites are the most abundant ones and with the most primitive elemental compositions; therefore, they can give us information about the composition and structure of the material in the primitive Solar System  \citep{beitz2016}. The volume filling factors measured for chondrites by \citet{Macke2011} are between $\phi = 0.58$ and $\phi = 1.0$. \citet{Beitz2013} performed impact experiments onto samples with compositions analogue to chondritic meteorites to investigate the dynamic-pressure range to achieve the degree of compaction found in these objects. They conclude that the impact pressures required for the observed compaction range between $\sim 0.05$ GPa and $\sim 2$ GPa.

The collisions outcome between pre-planetary bodies is determined by several physical parameters; one of them is the tensile strength. The tensile strength is the maximum tension that a material can resist before it breaks. This parameter is key for planet formation simulations that include models of planetesimal fragmentation 
\citep{SanSebastian2019}. The tensile strength range of meteoroid-stream materials of likely cometary origin is between 0.4 kPa and 150 kPa \citep{Blum2014}.  Tensile strengths for different types of meteorites cover a wide range of values from about 0.7 MPa to 403 MPa \citep{Ostrowski2019}. The tensile strength measured in ordinary chondrites is in the range from 18 to 31 MPa \citep{Slyuta2017}, and the tensile strength measured in stony meteorites is in the range from 2 to 56 MPa \citep{Flynn2018}. Several experimental studies measured the tensile strength of dust aggregates; in particular, \citet{Meisner2012} and \citet{Gundlach2018} utilized the Brazilian Disk Test to obtain the tensile strength of cylindrical centimetre-sized samples. \citet{Gundlach2018} measured mean tensile strengths of hand-compressed samples consisting of $\mathrm{\mu m}$-sized water-ice grains and found that they are comparable to the values found for $\mathrm{\mu m}$-sized silica grains and generally in good agreement with astrophysical observations, such as the COSIMA (Cometary Secondary Ion Mass Analyser) strength estimates for the dust of comet 67P Churyumov-Gerasimenko \citep{Hornung2016}. This suggests that icy aggregates are not harder to destroy than non-icy aggregates. The tensile strength derived by
\citet{Meisner2012} varies between 1 kPa and 6 kPa for cm-sized dust samples of volume filling factors between 0.34 and 0.50. The results found in both papers can be used as input parameters in detailed numerical simulations of collisions.

When a projectile of mass $M_p$ collides with a target of mass $M_t$ (with $M_t \gg M_p$) at a velocity $v_{col}$ and specific energy $Q$, the outcome of the impact may result in accretion, shattering or dispersal of the target. In particular, for the case of dispersal, the threshold for catastrophic disruption $Q_D^*$ is defined as the minimum specific energy needed to disperse the target in two or more pieces, with the largest fragment having a mass $M_t /2$. The set of physical properties defining  $Q_D^*$ is parameterized from impact experiments \citep{Bukhari2017,Beitz2011} and from smoothed-particle hydrodynamics (SPH) simulations \citep{BenzAsphaug1999,Benz2000}. \citet{BenzAsphaug1999} and \citet{Benz2000} obtained the specific energy for catastrophic disruption, $Q_D^*$ of solid bodies ($\phi= 1$) with SPH simulations for basaltic targets at $v_{col}$= 3 and 5 km/s \citep{BenzAsphaug1999} and 20-30 m/s \citep{Benz2000}. For weak targets (solid water ice), \citet{BenzAsphaug1999} also obtained $Q_D^*$ at $v_{col}$= 0.5 and 3 km/s. \citet{Jutzi2010} carried out similar SPH simulations, but including the calculation of $Q_D^*$ for pumice ($\phi = 0.1$). They found that $Q_D^*$ for porous targets is greater than for non-porous in the strength-dominated regime (SR), while it is a bit smaller in the gravity-dominated regime (GR). For all SPH simulation, the transition from the SR to the GR in $Q_D^*$ occurs for planetesimals of radius $\sim 100$ m. \citet{Beitz2011} and \citet{Bukhari2017} showed in low-velocity impact experiments that the strength of compacted dust aggregates is much weaker than that of porous rocks. Thus, the transition from the SR to the GR may occur for objects of only 0.1 m in radius, if these are primitive dust aggregates. This new outcome may have important implications for models of planetary formation. 

Several planetary formation models \citep[e.g.][]{Chambers2014,Guilera2014}, which calculate the formation of protoplanets, including fragmentation of planetesimals, use the values of $Q_D^*$ obtained by \citet{BenzAsphaug1999} for basalt colliding at impact velocities of 3 km/s. However, planetesimals have different compositions and collision velocities, depending, among other parameters, on their location in the protoplanetary disk. Also, as planetary embryos grow, the planetesimal velocity dispersion increases because of gravitational excitations produced by planetary embryos. The increasing relative velocities among planetesimals cause them to fragment through mutual collisions.  \citet{SanSebastian2019,Chambers2014,Guilera2014},  studied the growth of a giant planet by the accretion of non-porous planetesimals, taking into account their fragmentation through mutual collisions. Most models of planetary formation, including planetesimal fragmentation, use the values of $Q_D^*$ for non-porous bodies and for a fixed impact velocity, assuming a single planetesimal composition \citep[e.g.][]{Chambers2014,Guilera2014}. In our fragmentation model \citep{SanSebastian2019}, we included the dependence of $Q_D^*$ on the composition of the non-porous planetesimals assuming a mixture of rock and ice beyond the iceline. We also included the dependence of $Q_D^*$ on the  impact velocities of the planetesimals. We showed that these improvements  significantly change the evolution of the growing protoplanet. Thus, the inclusion of the porosity in the internal structure of the planetesimals and the study of the collisional evolution of the porosity would have an impact on models of planetary formation. Moreover, the analysis and modelling of the collisional evolution of bodies in planetary systems is important not only to study the planetary formation process but also to be able to interpret disk observations \citep{Krivov2018}.

In this paper, we produced compressed dust samples by high-speed impact experiments, following \citet{Beitz2013} and measured the degree of compaction we could achieve by this method. We studied the relation between the impact velocity and the volume filling factor of the compressed samples.
Later, we performed the Brazilian Disk Test with the compressed samples to obtain their tensile strengths. The aim of this work is to calculate the critical fragmentation strength of the compressed samples and the relation with their volume filling factors. In Section \ref{sec2}, we present the first experiment, which consists of the impact compression of the samples. In Section \ref{sec2-1}, the experimental method and in Section \ref{sec2-2}, the results of the impact-compression experiments are presented. In Section \ref{sec3}, we describe the second experiment, which is the Brazilian Disk Test of the pre-compressed samples. In Section \ref{sec3-1}, we explain the experimental technique and in Section  \ref{sec3-2}, the obtained results. In Section \ref{sec4}, we present the calculation of the critical fragmentation strength. Finally, in Section \ref{sec5}, we draw the conclusions of our work.

\section{First experiment: Sample compression}
\label{sec2}

We carried out impact experiments to dynamically compress cylindrical dust samples. The aim of this first experiment was to produce highly compressed samples \citep[following][]{Beitz2013}, which represent the volume filling factors measured in chondrites \citep{Macke2011}. In the next subsections, we describe the method and setup developed for this task and the results obtained.

\subsection{Experimental technique}
\label{sec2-1}

\begin{figure}
  \centering
       \includegraphics[width=\linewidth]{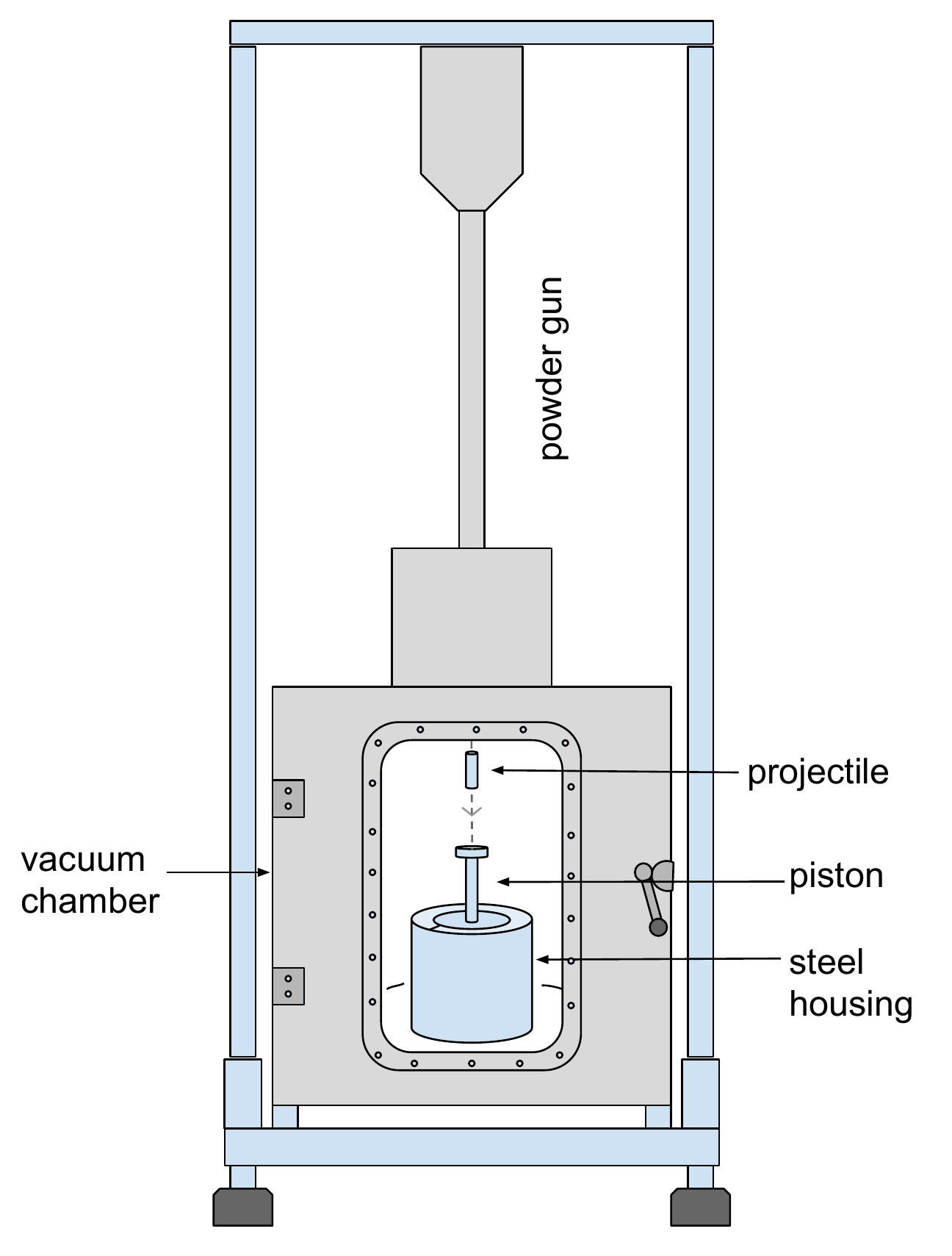}
        \caption{Schematic sketch of the experimental setup for the impact compression of dust samples. A powder gun accelerates a cylindrical projectile, which then impacts on a piston that compresses the dust sample inside a steel housing. Not shown is the high-speed camera, which records the impact through the front window of the vacuum chamber.}
  \label{fig:experimentalsetup1}
\end{figure}

\begin{figure}
 \centering
    \includegraphics[height=4.3cm, width=3.5cm, angle=-90 ]{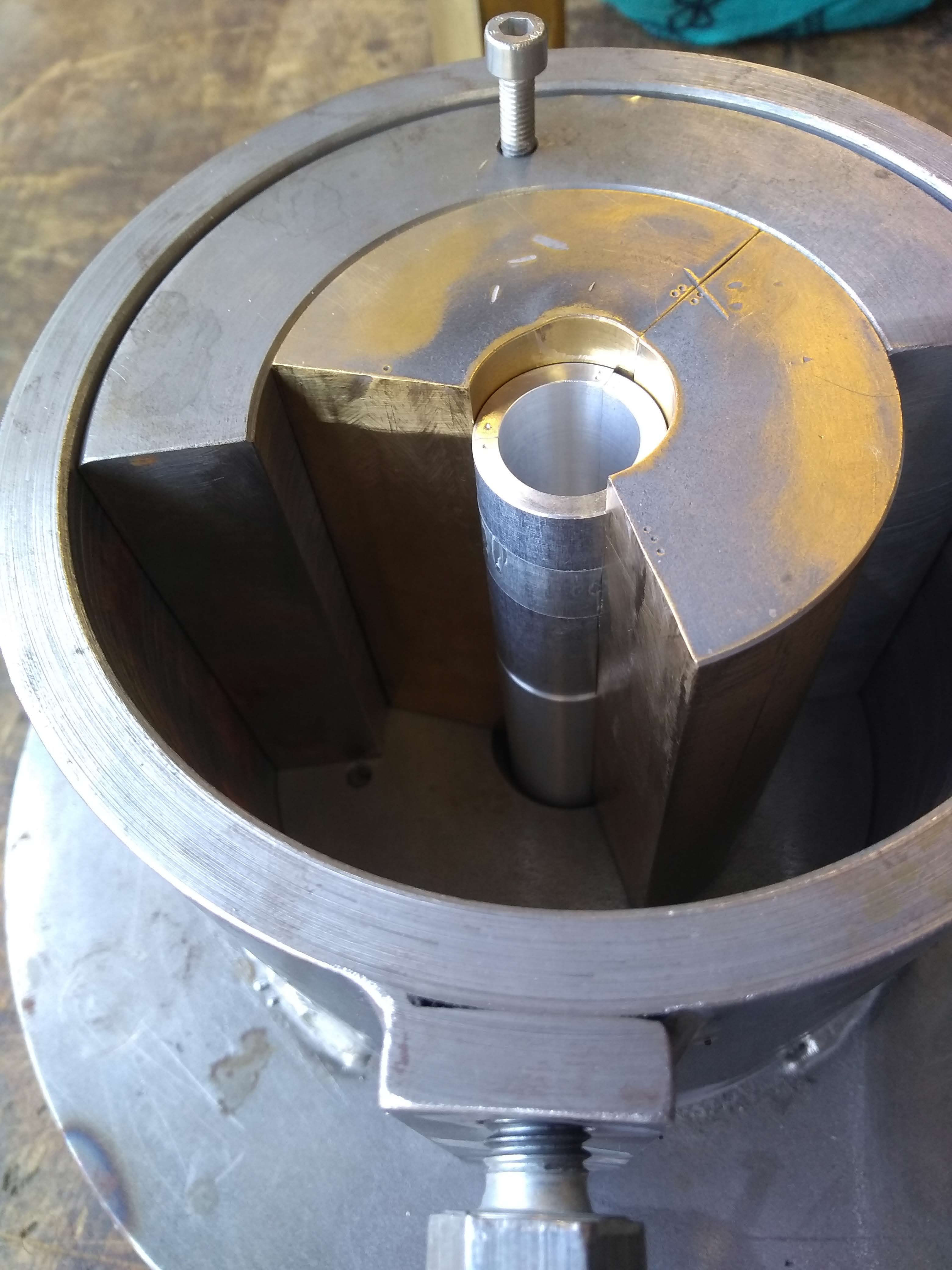} 
  \includegraphics[height=3.9cm, width=3.5cm, angle=-90]{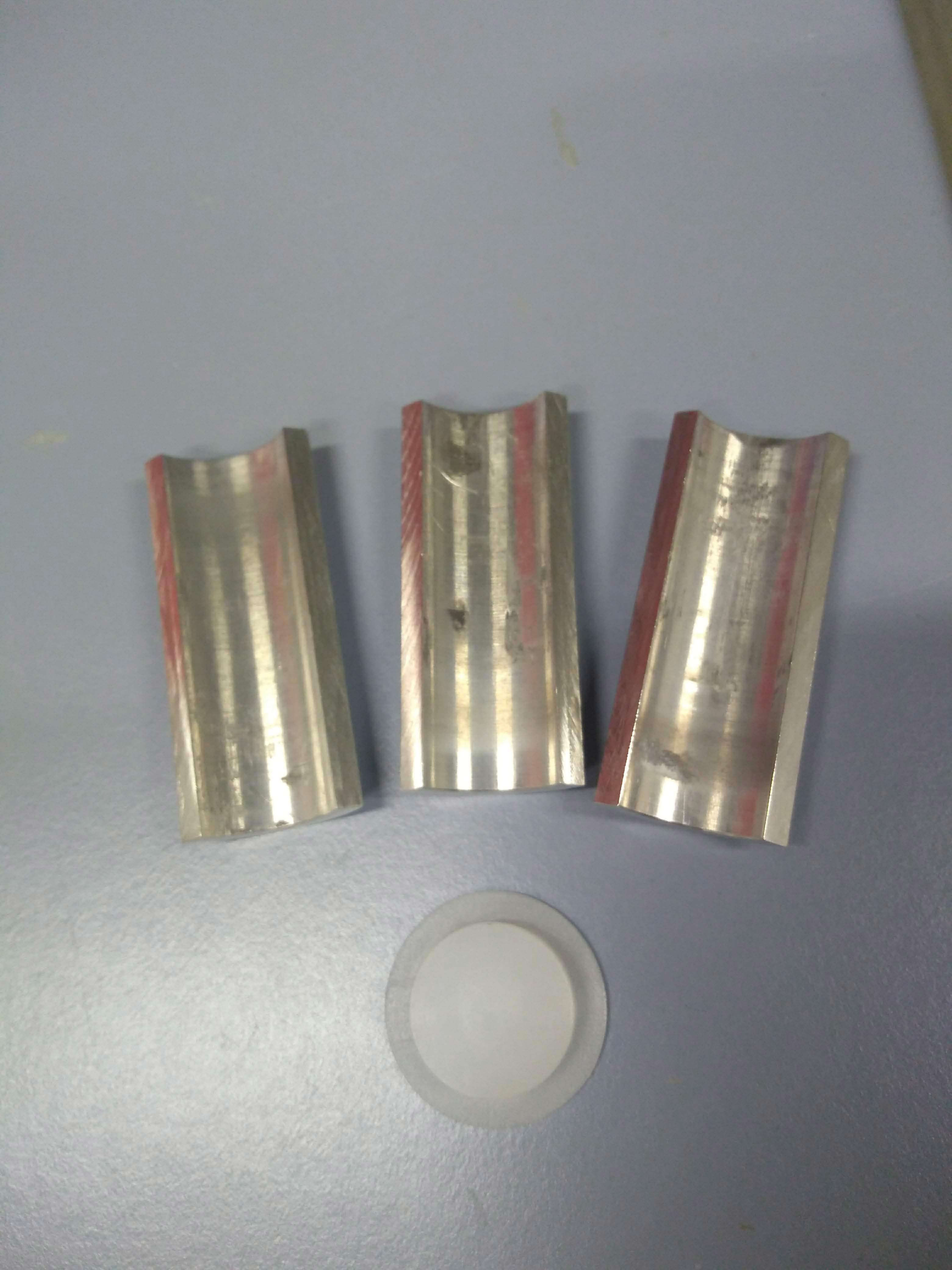}
  \caption{Left panel: Steel housing. Right panel: Aluminium holder with plastic base. The inner diameter of the holder is 1.6 cm.}
  \label{fig:experimentalsetup2}
\end{figure}


\begin{figure}
  \centering
    \includegraphics[width=0.32\linewidth]{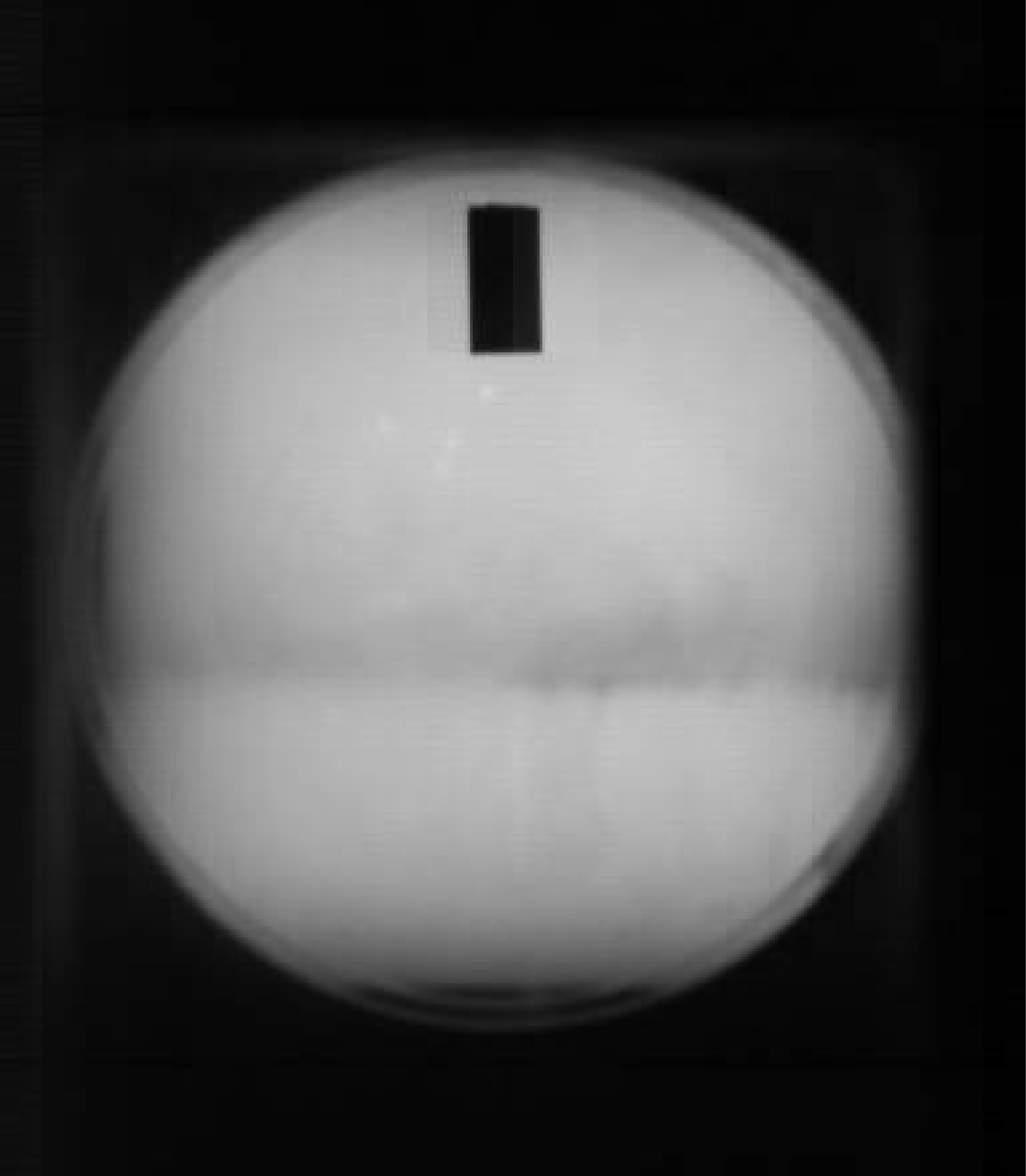} 
   \includegraphics[width=0.32\linewidth]{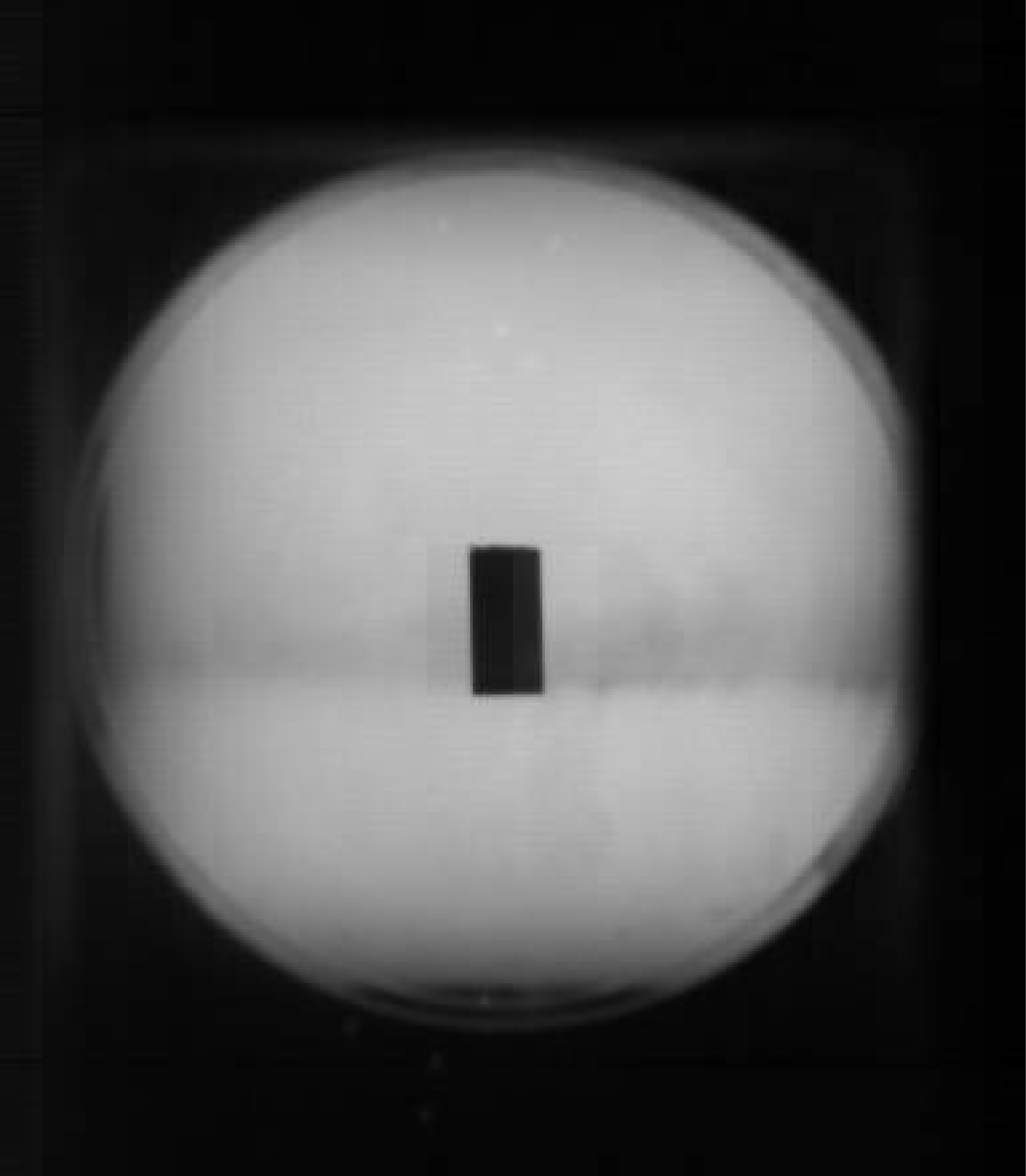} 
 \includegraphics[width=0.32\linewidth]{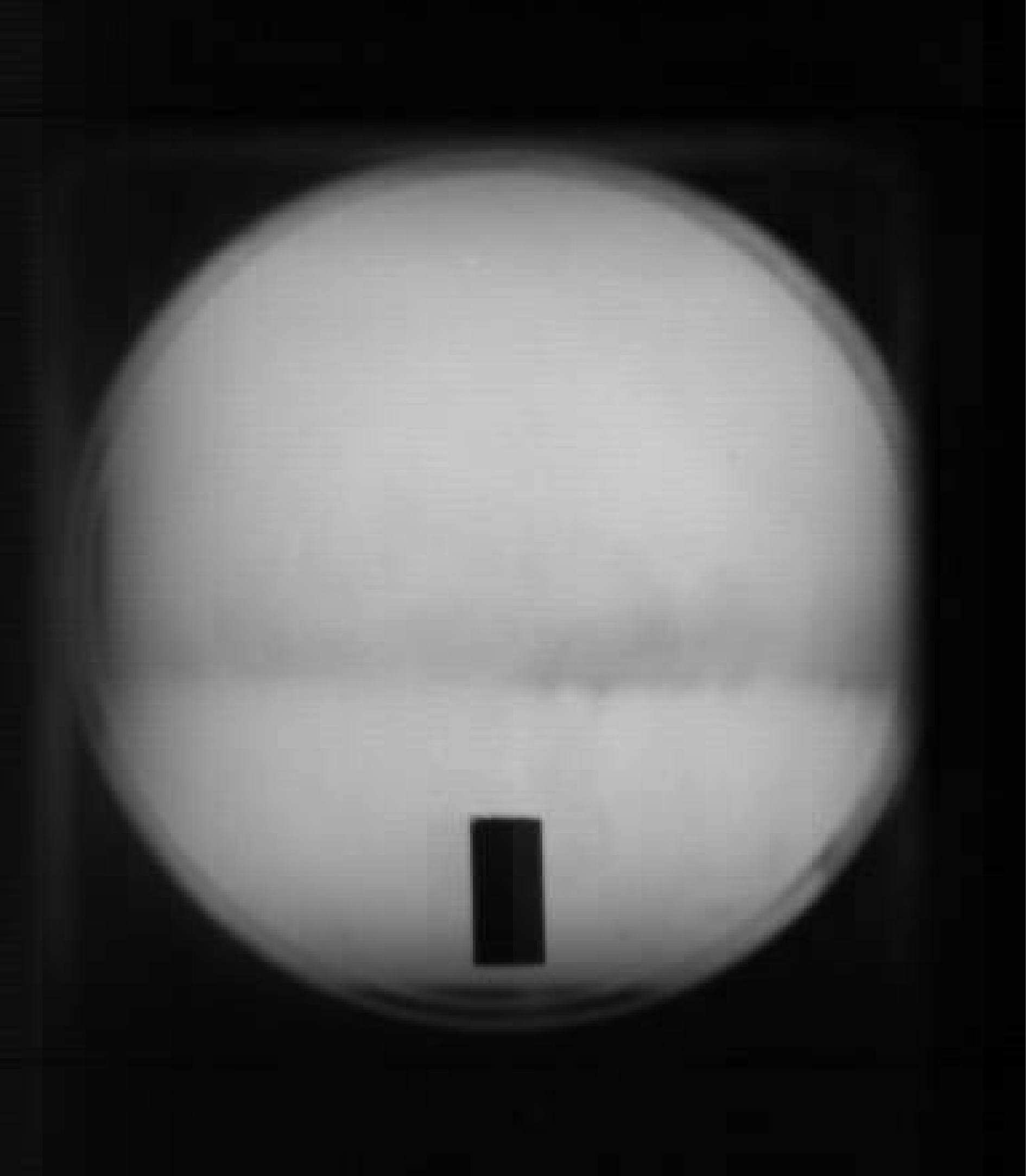} 
 \caption{Example of a projectile's trajectory recorded by the high-speed camera. Size of the cylindrical projectile: 1.5 cm diameter and 3 cm length. Impact velocity: 62 $\pm$ 4 m/s. Frame rate: 37,500 frames per second.}
 \label{fig:projectilesequence}
\end{figure}

The aim of this experiment was to produce compacted dust samples with 
volume filling factors that represent those measured in chondrites \citep{Macke2011}, i.e., $\phi=0.6-1$, by using a vertical powder gun at
IGeP, TU Braunschweig (see Fig. \ref{fig:experimentalsetup1}). We used micron-sized irregular silica grains with a size range between $\sim$0.5 and 10 $\mu$m (80$\%$ distributed between 1 and 5 $\mu$m) as an anolog
material of the chondritic meteorite composition as was done in
previous works \citep{Beitz2011,BlumWurm2008}.

The silica grains were filled into a three-part aluminium holder with a plastic base (see Fig. \ref{fig:experimentalsetup2}, right panel). This holder tube was enshrouded by a massive steel housing to prevent the tube from disassembling when being impacted by the projectile (see Fig. \ref{fig:experimentalsetup2}, left panel). The diameter of the dust samples  was 1.6 cm.

We included a piston on the top of the sample so that the projectile impacted there and not directly into the sample (see Fig. \ref{fig:experimentalsetup1}). We built the bottom part of the piston removable with two different materials; plastic or aluminium. Thus, if the bottom part was broken or deformed after the impact, we could exchange it. In general, for low impact velocities, we used the plastic pieces and for high impact velocities the aluminium ones. We also included a thin disk in between the sample and the piston, since the piston often adhered to the sample when the impact velocity was high. As with the bottom part of the piston, we used two different materials for the thin disks; plastic and aluminium. On the top part of the piston, we included a removable thick metal disk to protect the piston from the impact.

The trajectory of the projectile was recorded using a high-speed camera. By measuring the position of the projectile through image analysis, we derived its impact velocity (see Fig. \ref{fig:projectilesequence}).

After the impact, we extracted the discoidal sample from the metal holder, measured its thickness and mass and saved it for the second experiment. Since the diameter of all the samples was 1.6 cm, we calculated the volume of each sample from its measured thickness $l$. With the mass of the sample and its volume, we then obtained its bulk density $\rho$ and therefore its volume filling factor $\phi= \rho/\rho_{0}$, where $\rho_{0} =$ 2.6 g cm$^{-3}$.

\subsection{Results}
\label{sec2-2}

In total, we performed 40 impact experiments with impact velocities from $\sim$ 60 m/s to $\sim$ 500 m/s, from which 20 compressed samples were successfully retrieved. We considered an impact experiment successful if  after the impact we could extract the sample intact and safely store it without breaking it. In the other 20 experiments, the samples broke during the impact, afterwards, when we extracted the sample from the mold, or  during the measurement of the extracted sample. From the intact samples, we measured volume and mass and determined their bulk densities, and thus, volume filling factors.

\begin{figure}
    \centering
    \includegraphics[width=\linewidth]{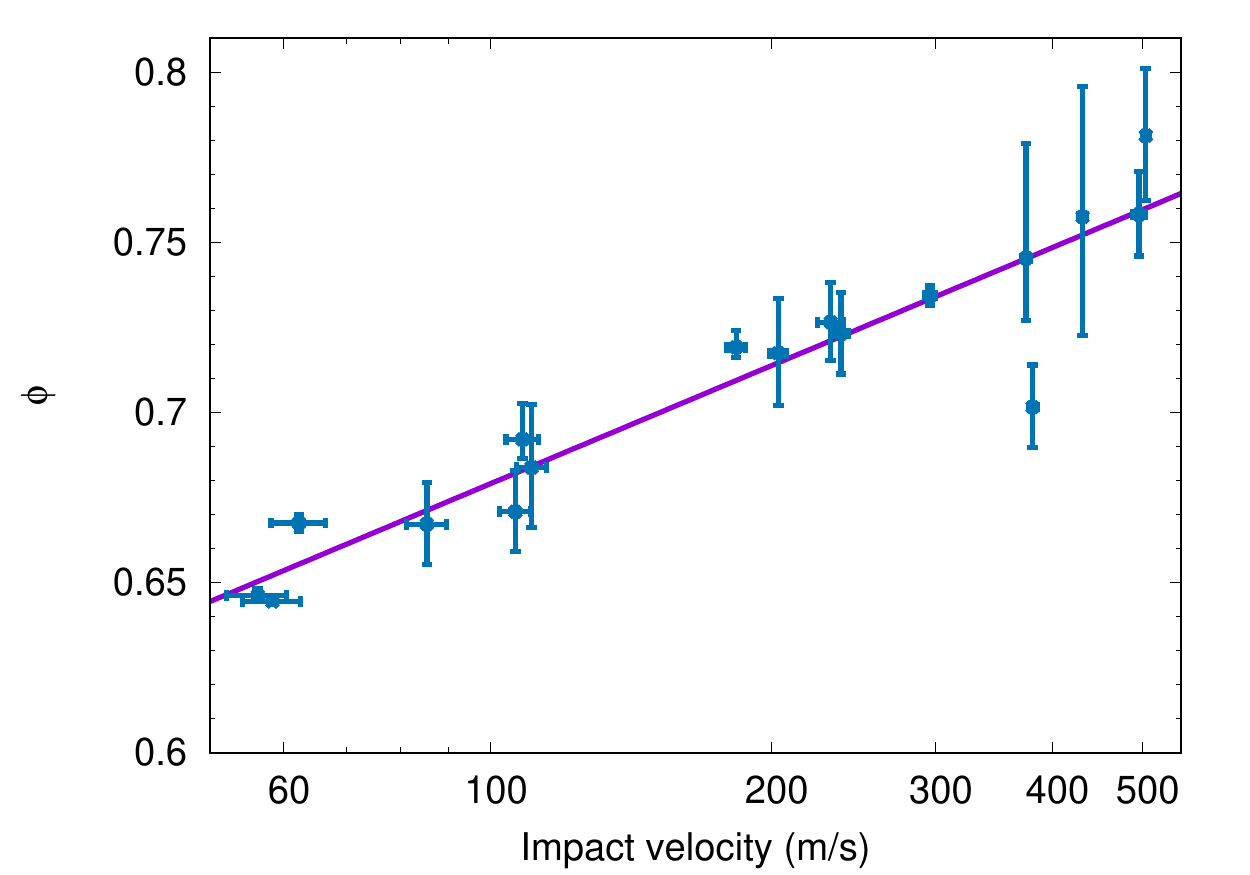}
    \caption{Volume filling factors of the impact-compressed samples as a function of the impact velocity of the projectile.}
    \label{fig:velvsvff}
\end{figure}

The results of the first experiment are presented in Fig. \ref{fig:velvsvff} where the solid line was added to guide the eye. We can see, as expected, that higher impact velocities result in higher volume filling factors.   

\citet{Beitz2013} performed impact experiments into samples of chondritic analogue material, achieving volume filling factors from 0.70 to 0.99. One of the aims of their work was to measure the degree of compaction of the samples to obtain a compaction law applicable to porous asteroids as the sources of chondritic meteorites \citep{beitz2016}. Their consolidated samples were recovered after the impacts and further analyzed by computer-aided X-ray tomography. Even though we followed their work in our experiments,
there are two major differences: (i) \citet{Beitz2013} left their samples in a plastic holder, while we intended to extract the samples from the holder to be able to apply later the Brazilian Disk Test, making it more challenging to achieve intact samples with high volume filling factors. (ii) \citet{Beitz2013} shot the projectile directly into the sample, while we had to incorporate a piston to protect our sample so that our impact pressure that could reach the sample at a given impact velocity presumably was lower.

For velocities higher than 500 m/s, we had an experimental limit since the setup deformed after the impact. In the future, we plan to develop a different setup that can endure impacts at higher velocities and therefore achieve samples with higher volume filling factors.

\section{Second experiment: Brazilian Disk Test}
\label{sec3}

The Brazilian Disk Test (BDT) is an engineering method to measure the tensile strength of a material. We performed the BDT for the compressed samples described in the previous Section to obtain the values of their tensile strength. In the following subsections, we describe the experimental technique, the setup used for the BDT and the results obtained with the experiment.

\subsection{Experimental technique}
\label{sec3-1}

\begin{figure}
    \centering
    \includegraphics[width=\linewidth]{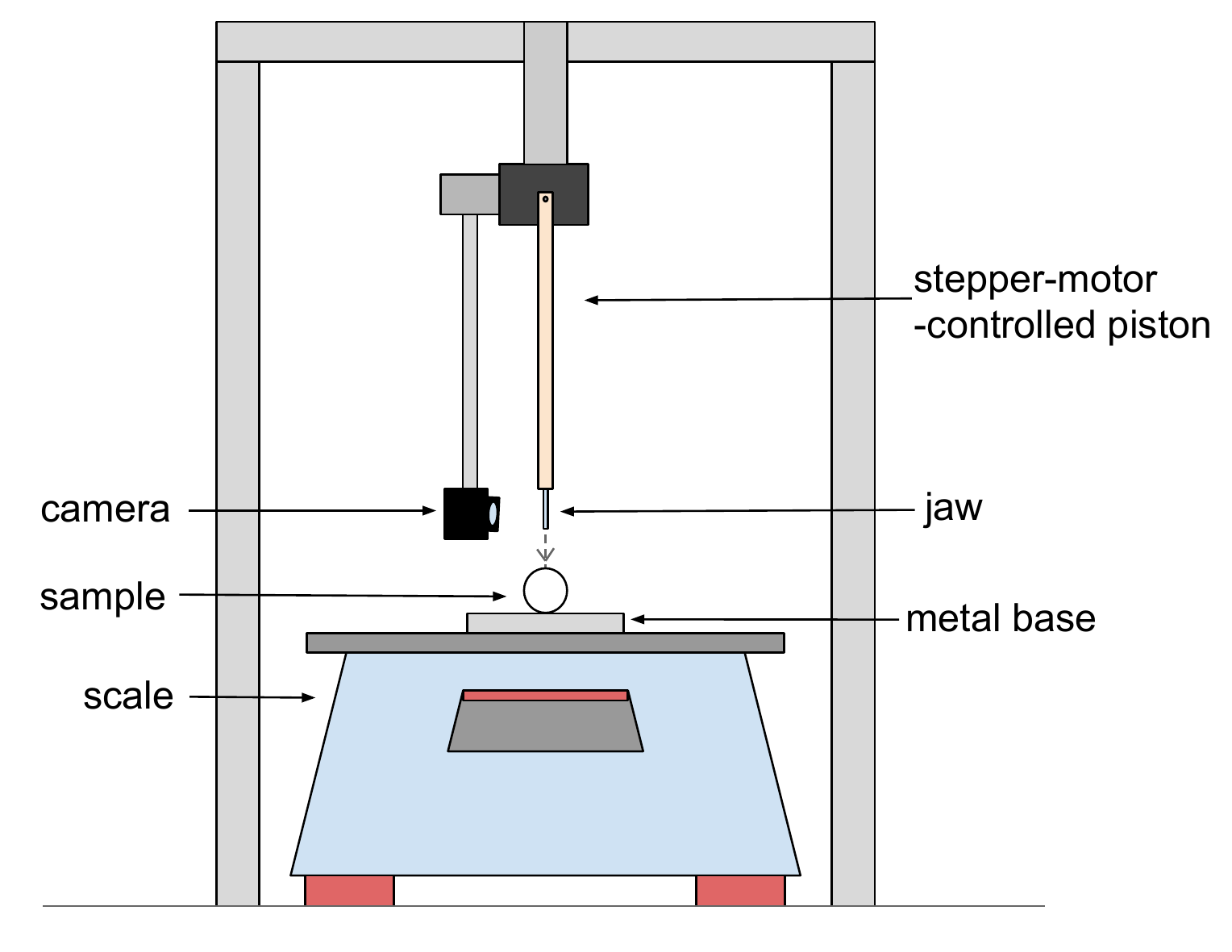}
    \caption{Schematic setup of the Brazilian Disk Test with its components.}
    \label{fig:bdtsetup}
\end{figure}

\begin{figure}
  \centering
     \includegraphics[width=0.45\linewidth]{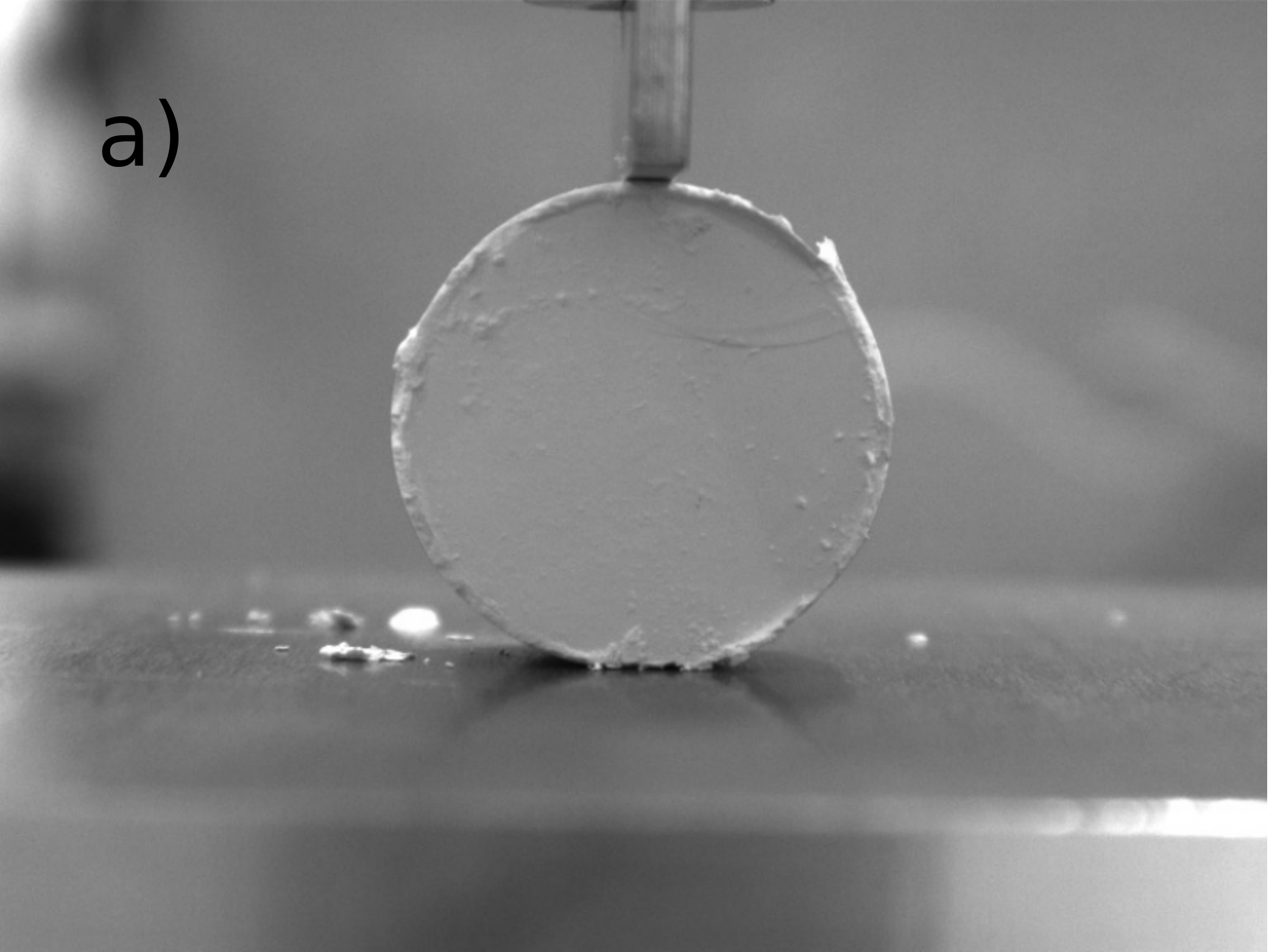}
        \includegraphics[width=0.45\linewidth]{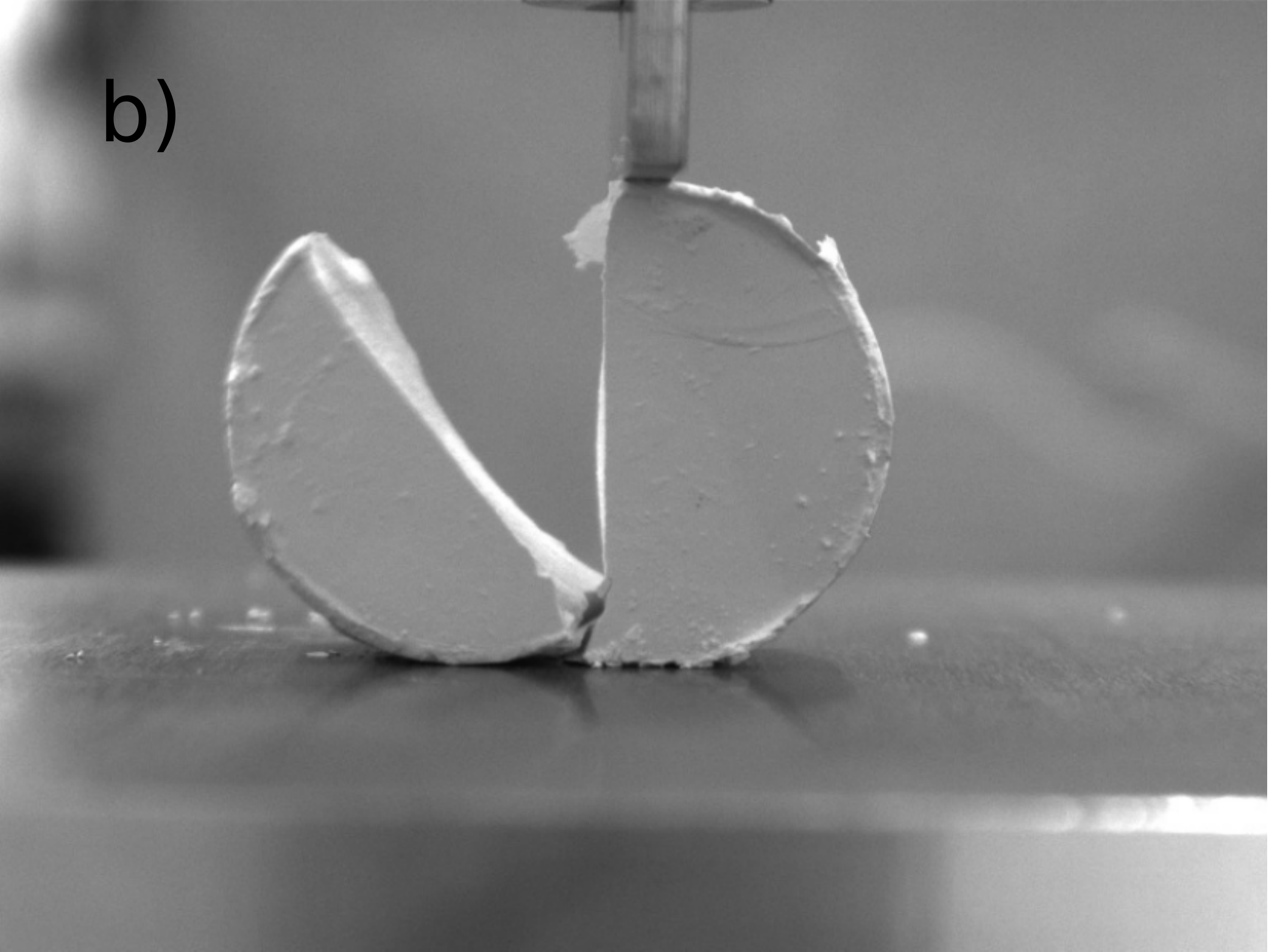}\\
         \includegraphics[width=0.95\linewidth]{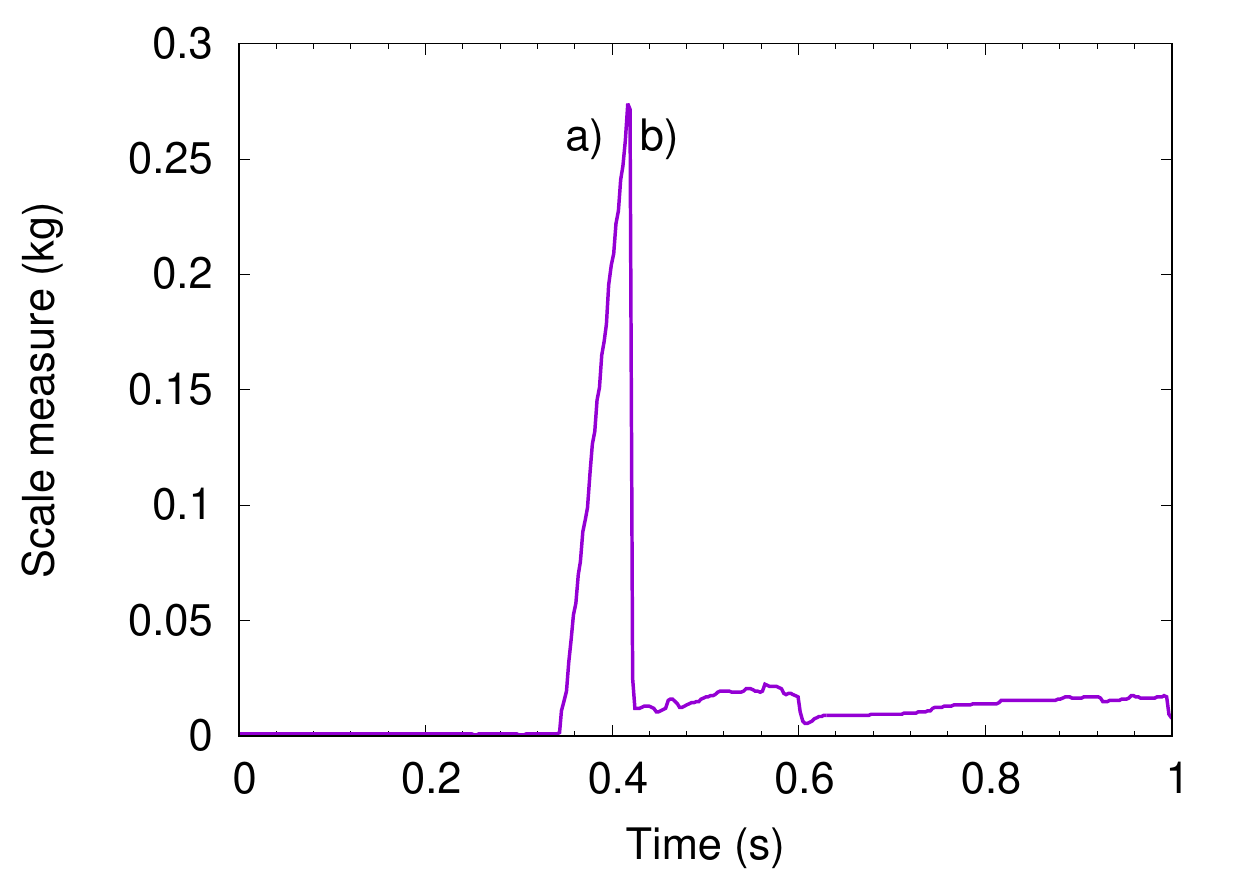}
  \caption{Top: Compact silica aggregate before (left) and after breakup (right). Bottom: Time evolution of the scale readings during a BDT experiment. Before the maximum corresponds to a) (top left) and after the maximum to b) (top right).}
  \label{fig:crack}
\end{figure}

The BDT \citep{LiWong2013} consists of compressing the pre-compacted circular dust disk until it breaks into two similar-sized pieces. One can estimate the tensile strength at the moment of the breakup by 
\begin{equation}
    \sigma = \frac{2 F}{\pi d l}, 
	\label{eq:tensilestrength}
\end{equation}
where $F$, $d$ and $l$ are the exerted force at breakup (measured by a balance), the diameter (which is 1.6 cm for all our samples), and the thickness of the dust cylinder, respectively \citep{Meisner2012,Gundlach2018}.

The configuration of the experiment is shown in Fig. \ref{fig:bdtsetup}. The result of the BDT can be influenced by the jaw's curvature, but from \citet{M&K2016} we know that the influence is less than $\sim$ 5 \% (see their Figure 12). In Fig. \ref{fig:crack}, we can observe the images of a pre-compacted silica aggregate taken by the camera before (a) and
after (b) the breakup, and the measurements taken by the scale during an experimental run.

\subsection{Results}
\label{sec3-2}

\begin{figure}
    \centering
    \includegraphics[width=\linewidth]{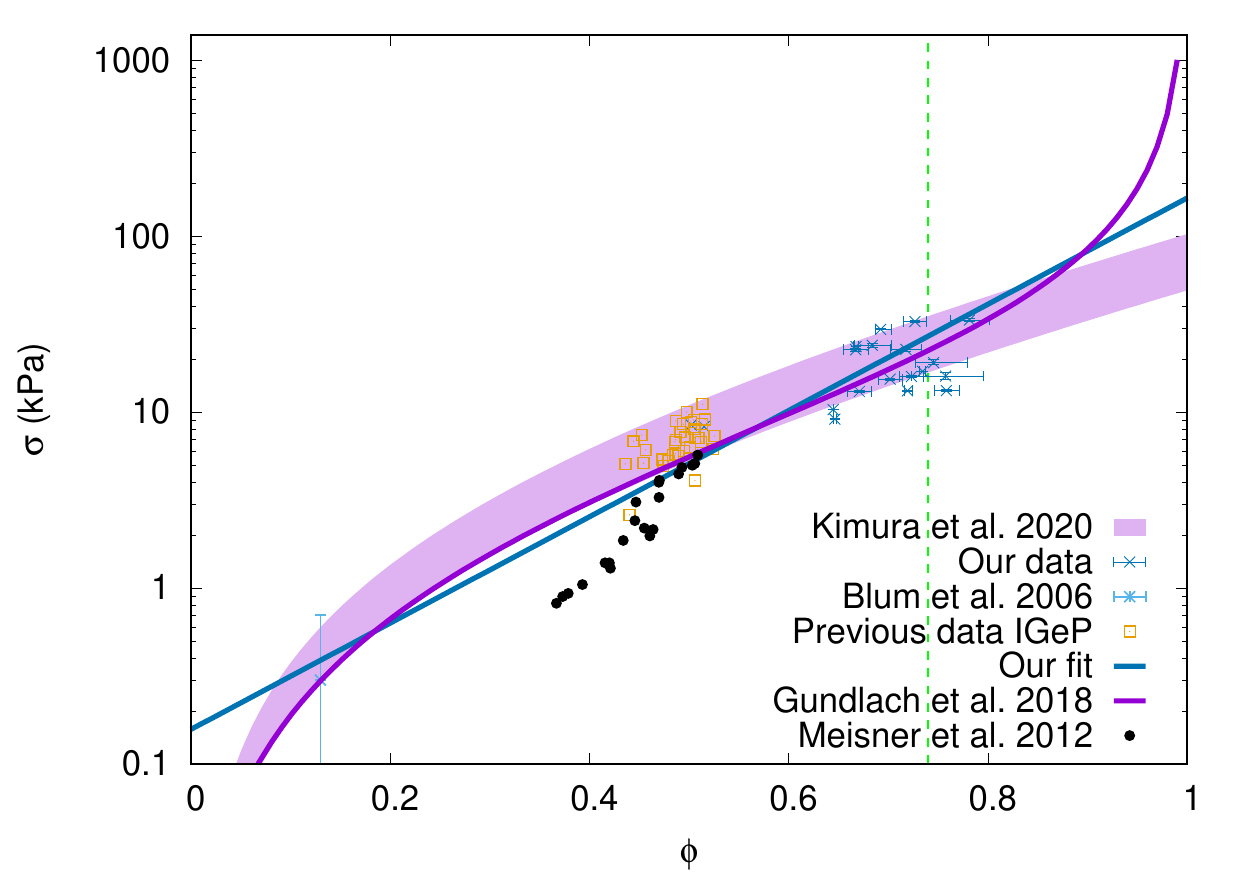}
    \caption{Dependence of the tensile strength on the volume filling factor. Our tensile-strength values (blue points) as well as those of \citet{Meisner2012} (black points) were obtained with the Brazilian Disk Test. Solid blue line: our fit including data extracted from \citet{Blum2006} (light blue point with error bar), data from previous experiments done at IGeP, TU Braunschweig (yellow points), data extracted from \citet{Meisner2012} and our data. Solid purple line: fit to the data adopting the function by \citet{Gundlach2018}. Purple area: fit to our data using the relation between tensile strength and volume filling factor by \citet{Kimura2020}. The green dashed line corresponds to $\phi=0.74$ (see text for details). }
    \label{fig:tvsvff}
\end{figure}

It is important to note that our samples did not have a perfectly uniform porosity. Due to the impact dynamics, the top and bottom circular layers of the disks had a slightly higher density than the disks' inner parts. In some of the BDT experiments, one or both of these layers did not break into two halves, in contrast to  the inner part that always broke. Therefore, our results have to be taken as a lower limit for the real tensile strength. Nevertheless, \citet{Beitz2013} showed that the porosity of impact-compressed samples varied by less than 10\% over their total height (see their figure 4). It should be noted that due to the material inhomogeneity, the crack initiation point might not always start at the centre but at the edges of the load contact \citep{LiWong2013}.

In  Fig. \ref{fig:tvsvff}, we can observe a positive correlation between the tensile strength, measured by the BDT, and the volume filling factor. Using also literature data \citep{Meisner2012,Blum2006}, unpublished measurements of samples compressed by hand, which were previously obtained at IGeP, TU Braunschweig, and our new results, we can follow the description by \cite{Meisner2012} of the tensile strength as a function of volume filling factor
\begin{equation}
    \sigma (\phi) = a \; e^{b \phi},
    	\label{eq:meisner}
\end{equation}
\noindent  where fitting through the least squares method, we obtain $a= 0.16 \pm 0.04$ and $b= 6.96 \pm 0.47$. The difference between the fitting parameters of \cite{Meisner2012} and our work is due to the inclusion of the much wider range of volume filling factors when our compressed samples are incorporated into the data set. 

In addition, we also fitted another functional behavior to the full data set, adopting the correction function given by \cite{Gundlach2018}, and obtained
\begin{equation}
    \sigma_{\mathrm{cor}}(\phi) = \phi*\left(\frac{c}{1-\phi}-d\right).
    	\label{eq:gundlach}
\end{equation}
\noindent We fit the data using the least squares method, giving a weight to each data point, which takes into account the number of data points in each set of experiments and the value of the tensile strength. With this, we give equal weight to all data sets and also compensate for the large variation in tensile-strength values. We obtain $c= 10.40 \pm 0.74$ and $d= 9.67 \pm 0.87$. In Fig. \ref{fig:tvsvff}, both fit functions are shown as solid curves.

We remark that in the wide range of volume filling factors considered in this work, $0.1 \lesssim \phi \lesssim 0.9$, the differences between both approximations are rather small.

In Fig. \ref{fig:tvsvff}, we also show the relation between the tensile strength and the volume filling factor given by \citet{Kimura2020}, where they computed the tensile strength $\sigma_{V}$ of a dust aggregate consisting of $N$ elastic spherical monomers of radius $r_0$ and surface energy $\gamma$, including a dependence of $\sigma_{V}$ on the volume $V$ of the aggregate (their Eq. 5)
\begin{equation}
\begin{split}
    \sigma_{V}(\phi) = & 8 \; \mathrm{kPa} \left( \frac{\gamma}{0.1 \; \mathrm{J\; m^{-2}} } \right) \left( \frac{r_0}{0.1 \; \mathrm{\mu m}} \right)^{3/m -1} \left( \frac{\phi}{0.1} \right)^{\beta-1/m} \\
    & \mathrm{exp}\left[ \alpha \left( \frac{\phi}{0.1} -1 \right) \right] \left( \frac{V}{686 \; \mathrm{\mu m^3}} \right)^{-1/m}
    \end{split}
    	\label{eq:kimura}
\end{equation}
\noindent where $\alpha=0.24$ and $\beta=1.5$. The parameter $m$ is commonly referred to as the Weibull modulus, whose value is 8 for siliceous material, and $\gamma$ is the surface energy. Since the volume of our samples vary, we write it as a function of $\phi$, $V= \tilde{m}/(\phi \rho_0)$. $\tilde{m}$ is an average of the mass of our compressed samples with $\phi < 0.74$, because Eq.\ref{eq:kimura} can be applied, in principle, for volume filling factors lower than that value. Following \cite{Kothe2013}, we adopt two values for $r_0$ since the average particle size by number is 0.63 $\mathrm{\mu m}$  and by mass 2.05 $\mathrm{\mu m}$ for the irregular silica that we used in our samples. We fit Eq. \ref{eq:kimura} to our data, obtaining a value of the surface energy of $\gamma = 0.18 \pm 0.02 \, \mathrm{J\; m^{-2}}$ for $r_0 = 0.67 \mathrm{\mu m}$ (the mean value of the two values of $r_0$ adopted). Given this value of $\gamma$, we can conclude that absorbed water on the surface is not so important for our highly-compacted samples. This phenomenon was already noticed by \citet{Kimura2020}. As shown in Fig. \ref{fig:tvsvff}, in the range of $0.315 \mathrm{\mu m} <r_0< 1.025 \mathrm{\mu m}$, Eq. \ref{eq:kimura} can reproduce the values of the tensile strengths of our compressed samples consisting of irregular silica particles reasonably well.

It is important to remark that our results  of the tensile strength are in agreement with the range given by \cite{Blum2014}.

\section{Critical fragmentation strength and catastrophic disruption threshold}
\label{sec4}

\begin{figure}
    \centering
   \includegraphics[width=\linewidth]{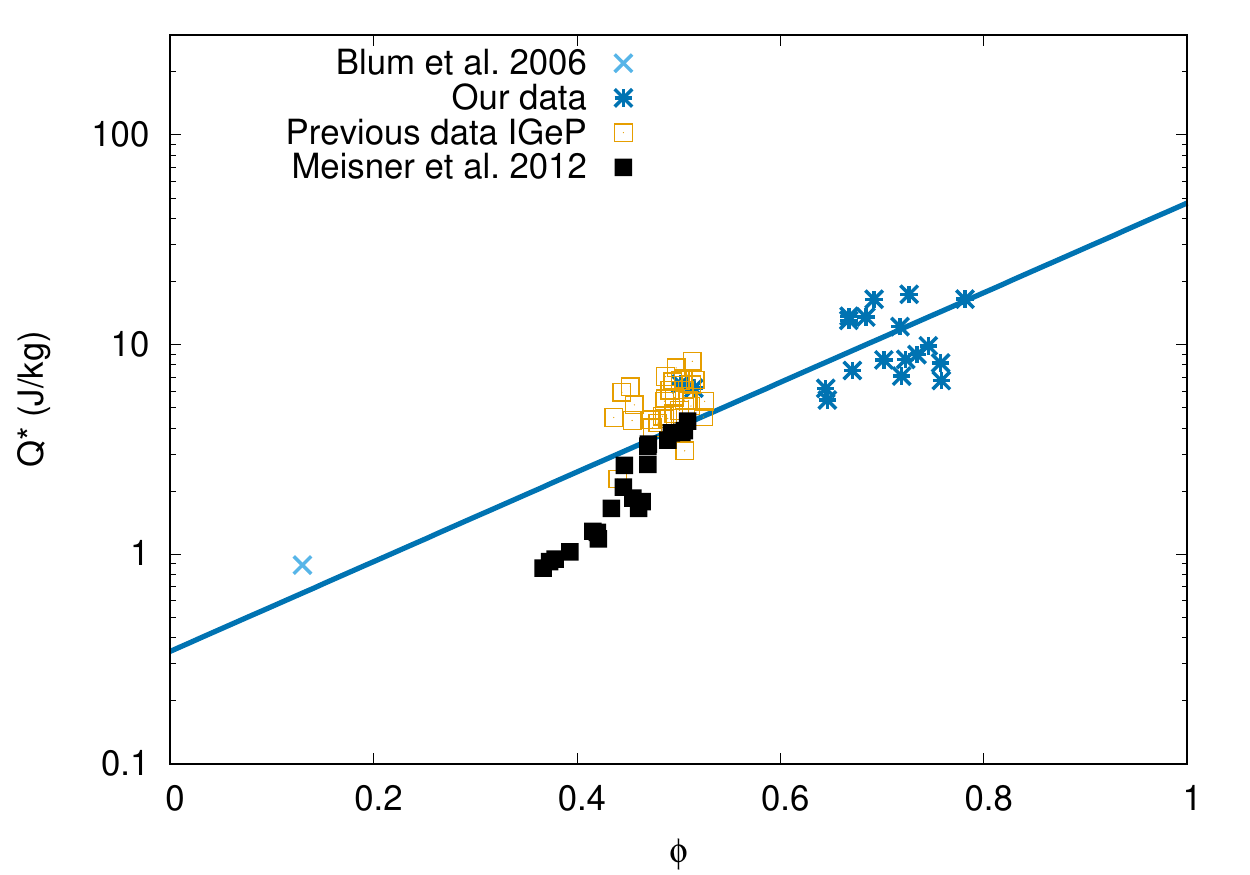}
    \caption{Critical fragmentation strength as a function of volume filling factor. The solid line is an exponential fit to the data.}
    \label{fig:qd*}
    \end{figure}

In  the previous Section, we derived the tensile strength of dynamically compressed dust samples using the BDT. As we mentioned before, the tensile strength $\sigma$ is the maximum pressure a material can resist before it breaks. Therefore, we can relate the
tensile strength to the dynamical pressure at the moment of impact in a collision, i.e.,
\begin{equation}
    \sigma = p = \frac{E_{\mathrm{kin}}}{V_{\mathrm{target}}},
    \label{eq:pressure}
\end{equation}
\noindent where $p$ is the impact pressure, $E_{\mathrm{kin}}$ the kinetic energy of the projectile and $V_{\mathrm{target}}$ the target volume. From Eq. \ref{eq:pressure}, we can obtain the impact velocity  that a projectile would need to break the target. Also, we can relate the impact pressure $p$ to the critical fragmentation strength, which is the energy per unit mass required to break a body into fragments with the largest one having half the mass of the initial target \citep{BenzAsphaug1999}. Since with the BDT we obtained the pressure at which the samples are broken into two pieces of almost identical size -- which means that after the breakup we have the largest fragment with half the mass of the initial target -- we can relate this pressure with the critical fragmentation strength $Q^*$.

Thus, assuming that the projectile has the same porosity as the target and dividing Eq. \ref{eq:pressure} by the bulk density, we can obtain  the critical fragmentation strength
\begin{equation}
    Q^* = \frac{\sigma}{\rho},
    \label{eq:fragenergy}
\end{equation}
\noindent where  $\rho  =  \rho_0  \phi$   and  $\rho_0  =  2.6$  g/cm$^3$ for $\mathrm{SiO_2}$. After calculating the values of $Q^*$ for all samples, we propose an exponential dependence $Q^* (\phi)=f \; \mathrm{exp}\left(g  \phi\right)$, where fitting through the least squares method, we obtain $f= 0.34 \pm 0.08$ and $g= 4.92 \pm 0.46$.

In Fig. \ref{fig:qd*}, we show the results of $Q^*$ as a function of the volume filling factor. With this relation, we can calculate $Q^*$ for any volume filling factor $\phi$. 

\begin{figure*}
    \centering
    \includegraphics[width=0.49\linewidth]{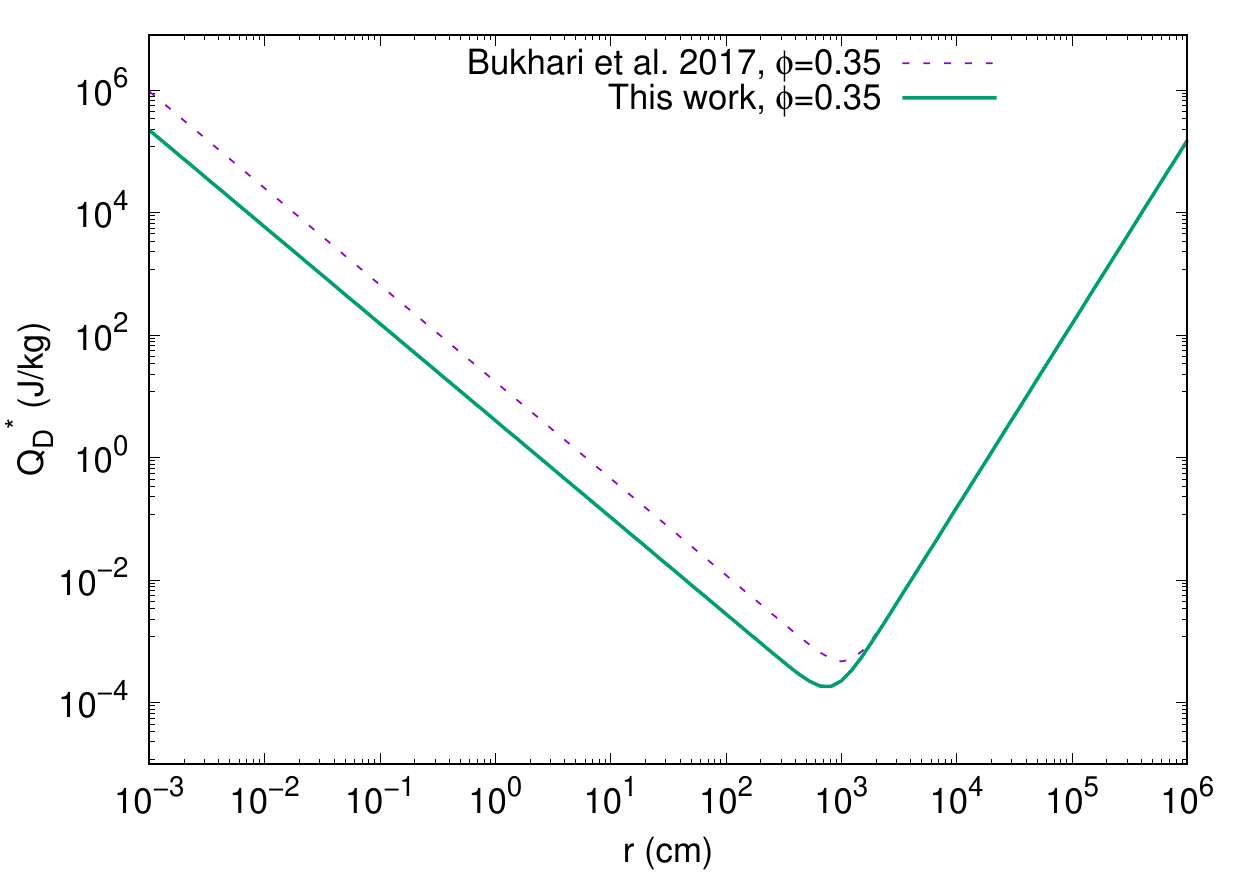}
     \includegraphics[width=0.49\linewidth]{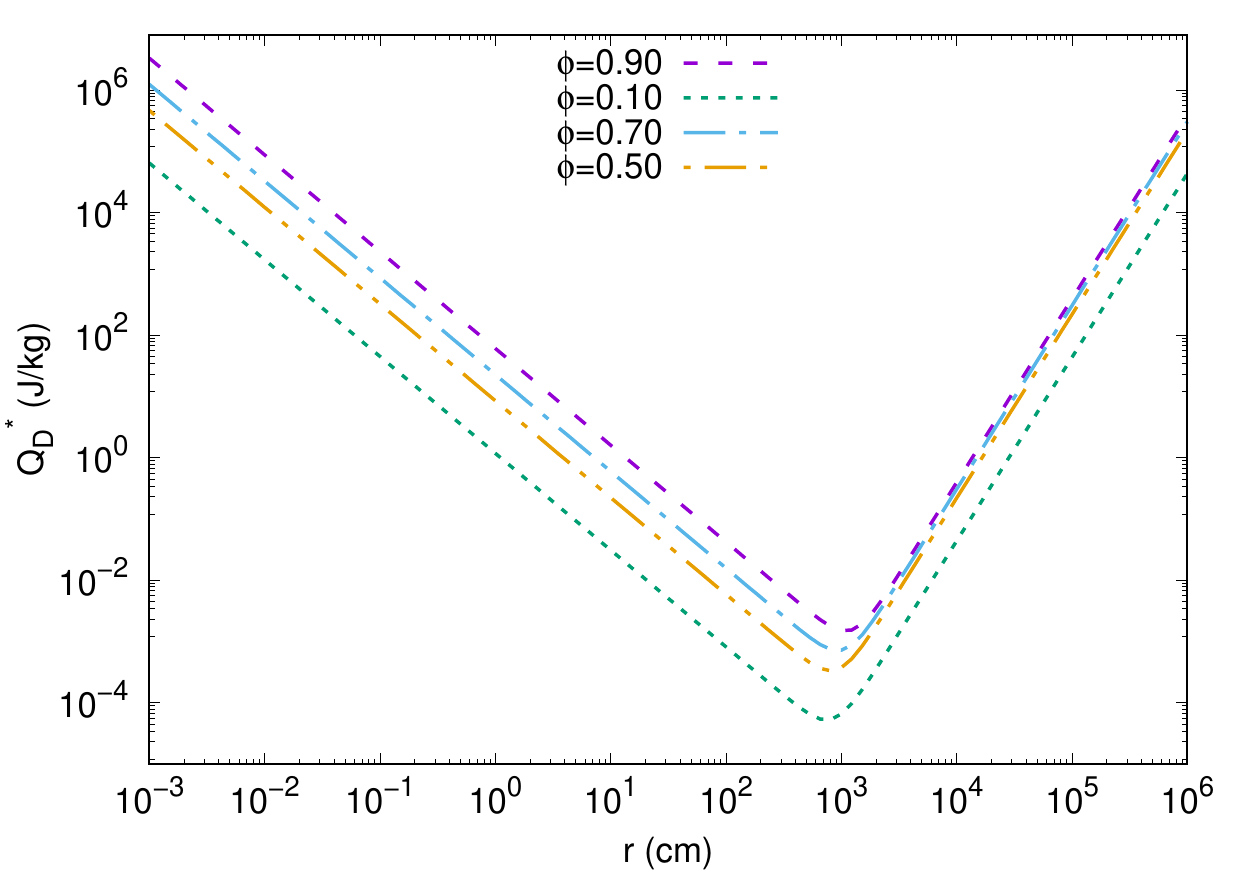}
    \caption{Catastrophic disruption threshold as a function of size of the colliding bodies. Left panel: comparison between the prescription given by \citet{Bukhari2017} and our prescription for a volume filling factor of 0.35. Right panel: our prescriptions for different volume filling factors.}
    \label{fig:qdcomparison}
    \end{figure*}

In a collision between two objects, the catastrophic disruption threshold required to fragment and disperse fifty per cent of the target mass (also known as specific impact energy or catastrophic impact energy threshold, $Q_D^*$) is an important property that is often used in collisional models. The catastrophic disruption threshold is given by $Q_D^{*} = Q^* + Q^*_l$ 
\citep{Armitage2010}, where $Q^*$ is the critical fragmentation strength and $Q^*_l$ is the specific energy required to gravitationally unbind the fragments containing the other 50 per cent of the target mass. The catastrophic impact energy threshold depends on many factors of the collision, in particular on the porosity of the colliding objects \citep{Jutzi2015}.

In this work, we propose an expression of $Q_D^{*}$ as a function of size for different values of the volume filling factor, which may be useful to be implemented in collisional models. We adopt the form of $Q_D^*$ obtained in \citet{Bukhari2017} (their Eq. 10), for samples consisting of micron.sized, irregular silica particles and a volume filling factor of 0.35, to fit our results and find the dependence of $Q^*$ on the volume filling factor in the strength regime SR. Here, we assume that the target and the projectile have the same size. Besides, following \citet{Krivov2018}, we add the term $Q^*_l$ corresponding to the gravitational regime GR, obtaining
\begin{equation}
    {Q_D^*} (r,\phi) = A(\phi) \left( \frac{r}{1 \; \mathrm{cm}}\right)^{-1.58} + \frac{4}{5} G \pi \left( \frac{r}{100 \; \mathrm{cm}} \right)^2 \rho_0 \phi, 
 \label{eq:final}
\end{equation}

\noindent where $r$ is the size of the target/projectile, $G$ the gravitational constant, $\rho_0  =  2600$  kg/m$^3$ and $A(\phi)$ is the parameter we fit for each value of the volume filling factor (Table \ref{tab:tableqd}).

 \begin{table}
 	\centering
 	\caption{Prescriptions of the catastrophic disruption threshold for different volume filling factors.}
 	\label{tab:tableqd}
 	\begin{tabular}{cc} 
 		\hline
 		$\phi$ & $A(\phi)$ \\
 	          &  (J/kg)   \\  
 		\hline
 		0.10 & 1.18  \\ 
 		0.35 &  4.05 \\
 		0.50 &  8.49 \\
 		0.70 & 22.72  \\
 		0.90 & 60.85  \\
 		\hline
 	\end{tabular}
 \end{table}

The results of Eq. \ref{eq:final} are shown in Fig. \ref{fig:qdcomparison} where we present the curves of $Q_D^*$ as a function of size, for $\phi=0.35$ compared with the prescription given by \citep{Bukhari2017} (left panel),   and for different volume filling factors (right panel).

\section{Conclusions}
\label{sec5}

In this work, we performed impact experiments to study the fragmentation of compact discoidal dust aggregates. The size of the dust samples was 1.6 cm of diameter, and we achieved compact samples with volume filling factors from about 0.64 to 0.8. In addition, we made compact dust samples compressed by hand with volume filling factors around 0.5. With these compact dust samples, we performed the Brazilian Disk Test to derive their tensile strengths. Following \citet{Meisner2012} and \citet{Gundlach2018}, we determined two correlations between the tensile strength and the volume filling factor for a wide range of this last parameter.

We can observe from Fig. \ref{fig:tvsvff} that the tensile strength values obtained in this work are in agreement with the strength values of meteoroid materials of cometary origin \citep{Blum2014}. On the other hand, the tensile strengths measured in all meteorites are of the order of megapascals \citep{Slyuta2017,Flynn2018,Ostrowski2019} and, as can be seen in Fig. \ref{fig:tvsvff}, these results can only be predicted by our Eq. \ref{eq:gundlach} for high volume filling factors $\phi \approx 1$. However, the tensile strength of asteroids is still a matter of debate. \citet{Grott2019} derived the tensile strength of a boulder on the C-type
asteroid Ryugu and obtained values on the order of a few hundred kilopascals. \citet{Grott2019} argue that there is an observational bias in our meteorite collections and the low tensile strength they predict indicates that any hypothetical meteoroid originating from the boulder would be too frail to survive the atmospheric entry. Meteorite porosities are fundamental parameters when studying the early Solar System and its formation process. The porosity affects most physical properties of meteorites, as \citet{Flynn2018} remark in their review, models of cratering and impact disruption may need to take this into account \citep{beitz2016}. 

It is important to note that our calculated relations depend on the size of the used particles. Following the upper and lower theoretical limits for the tensile strength of granular materials in \citet{Blum2006}, \citet{Gundlach2018} found an empirical formula for the grain size dependence of the tensile strength (their Eq. 4). Moreover, \citet{Kimura2020} present an analytical formula of the tensile strength as a function of the volume filling factor that incorporates the volume effect on the tensile strength. Applying their Eq. 5 for irregular silica particles, we found that it can reproduce the values of the tensile strength of our compressed samples (see Fig. \ref{fig:tvsvff}).

Finally, we  found a  relation  between  the critical  fragmentation strength and the volume filling factor. A better understanding of the relation between collisions and material porosities is key to understanding planetary formation processes. We provide new prescriptions for the catastrophic disruption threshold for a wide range of volume filling factors that can be applied 
to collisional models.

\section*{Acknowledgements}

This work was funded through a Short Research Project of the Deutscher Akademischer Austauschdienst (DAAD). M.G.P. thanks the DFG Research Unit ''Debris Disks in Planetary Systems'' for funding  her stay in Braunschweig. This work is part of the DFG project 298 BL/24-1. We are thankful to Kathrin Gebauer for her technical assistance in the experimental setup and to Christopher Kreuzig for the developed of the setup of the Brazilian Disk Test. We thank Tobias Eckhardt for providing us with his tensile-strength data for slightly compressed samples. We also thank the anonymous reviewer for helpful comments to improve the paper.



\section*{Data availability}

Data appearing in the figures are available upon request.



\bibliography{biblio} 

\begin{thebibliography}{}
\makeatletter
\relax
\def\mn@urlcharsother{\let\do\@makeother \do\$\do\&\do\#\do\^\do\_\do\%\do\~}
\def\mn@doi{\begingroup\mn@urlcharsother \@ifnextchar [ {\mn@doi@}
  {\mn@doi@[]}}
\def\mn@doi@[#1]#2{\def\@tempa{#1}\ifx\@tempa\@empty \href
  {http://dx.doi.org/#2} {doi:#2}\else \href {http://dx.doi.org/#2} {#1}\fi
  \endgroup}
\def\mn@eprint#1#2{\mn@eprint@#1:#2::\@nil}
\def\mn@eprint@arXiv#1{\href {http://arxiv.org/abs/#1} {{\tt arXiv:#1}}}
\def\mn@eprint@dblp#1{\href {http://dblp.uni-trier.de/rec/bibtex/#1.xml}
  {dblp:#1}}
\def\mn@eprint@#1:#2:#3:#4\@nil{\def\@tempa {#1}\def\@tempb {#2}\def\@tempc
  {#3}\ifx \@tempc \@empty \let \@tempc \@tempb \let \@tempb \@tempa \fi \ifx
  \@tempb \@empty \def\@tempb {arXiv}\fi \@ifundefined
  {mn@eprint@\@tempb}{\@tempb:\@tempc}{\expandafter \expandafter \csname
  mn@eprint@\@tempb\endcsname \expandafter{\@tempc}}}

\bibitem[\protect\citeauthoryear{{Armitage}}{{Armitage}}{2010}]{Armitage2010}
{Armitage} P.~J.,  2010, {Astrophysics of Planet Formation}

\bibitem[\protect\citeauthoryear{{Beitz}, {G{\"u}ttler}, {Blum}, {Meisner},
  {Teiser}  \& {Wurm}}{{Beitz} et~al.}{2011}]{Beitz2011}
{Beitz} E.,  {G{\"u}ttler} C.,  {Blum} J.,  {Meisner} T.,  {Teiser} J.,
  {Wurm} G.,  2011, \mn@doi [\apj] {10.1088/0004-637X/736/1/34}, \href
  {http://adsabs.harvard.edu/abs/2011ApJ...736...34B} {736, 34}

\bibitem[\protect\citeauthoryear{{Beitz}, {G{\"u}ttler}, {Nakamura},
  {Tsuchiyama}  \& {Blum}}{{Beitz} et~al.}{2013}]{Beitz2013}
{Beitz} E.,  {G{\"u}ttler} C.,  {Nakamura} A.~M.,  {Tsuchiyama} A.,   {Blum}
  J.,  2013, \mn@doi [\icarus] {10.1016/j.icarus.2013.04.028}, \href
  {http://adsabs.harvard.edu/abs/2013Icar..225..558B} {225, 558}

\bibitem[\protect\citeauthoryear{{Beitz}, {Blum}, {Parisi}  \&
  {Trigo-Rodriguez}}{{Beitz} et~al.}{2016}]{beitz2016}
{Beitz} E.,  {Blum} J.,  {Parisi} M.~G.,   {Trigo-Rodriguez} J.,  2016, \mn@doi
  [\apj] {10.3847/0004-637X/824/1/12}, \href
  {https://ui.adsabs.harvard.edu/abs/2016ApJ...824...12B} {824, 12}

\bibitem[\protect\citeauthoryear{{Benz}}{{Benz}}{2000}]{Benz2000}
{Benz} W.,  2000, \mn@doi [\ssr] {10.1023/A:1005207631229}, \href
  {http://adsabs.harvard.edu/abs/2000SSRv...92..279B} {92, 279}

\bibitem[\protect\citeauthoryear{{Benz} \& {Asphaug}}{{Benz} \&
  {Asphaug}}{1999}]{BenzAsphaug1999}
{Benz} W.,  {Asphaug} E.,  1999, \mn@doi [\icarus] {10.1006/icar.1999.6204},
  \href {http://adsabs.harvard.edu/abs/1999Icar..142....5B} {142, 5}

\bibitem[\protect\citeauthoryear{{Blum}}{{Blum}}{2018}]{blum2018}
{Blum} J.,  2018, \mn@doi [\ssr] {10.1007/s11214-018-0486-5}, \href
  {https://ui.adsabs.harvard.edu/abs/2018SSRv..214...52B} {214, 52}

\bibitem[\protect\citeauthoryear{{Blum} \& {Wurm}}{{Blum} \&
  {Wurm}}{2008}]{BlumWurm2008}
{Blum} J.,  {Wurm} G.,  2008, \mn@doi [\araa]
  {10.1146/annurev.astro.46.060407.145152}, \href
  {http://adsabs.harvard.edu/abs/2008ARA\%26A..46...21B} {46, 21}

\bibitem[\protect\citeauthoryear{{Blum}, {Schr{\"a}pler}, {Davidsson}  \&
  {Trigo-Rodr{\'{\i}}guez}}{{Blum} et~al.}{2006}]{Blum2006}
{Blum} J.,  {Schr{\"a}pler} R.,  {Davidsson} B.~J.~R.,
  {Trigo-Rodr{\'{\i}}guez} J.~M.,  2006, \mn@doi [\apj] {10.1086/508017}, \href
  {http://adsabs.harvard.edu/abs/2006ApJ...652.1768B} {652, 1768}

\bibitem[\protect\citeauthoryear{{Blum}, {Gundlach}, {M{\"u}hle}  \&
  {Trigo-Rodriguez}}{{Blum} et~al.}{2014}]{Blum2014}
{Blum} J.,  {Gundlach} B.,  {M{\"u}hle} S.,   {Trigo-Rodriguez} J.~M.,  2014,
  \mn@doi [\icarus] {10.1016/j.icarus.2014.03.016}, \href
  {http://adsabs.harvard.edu/abs/2014Icar..235..156B} {235, 156}

\bibitem[\protect\citeauthoryear{{Bukhari Syed}, {Blum}, {Wahlberg Jansson}  \&
  {Johansen}}{{Bukhari Syed} et~al.}{2017}]{Bukhari2017}
{Bukhari Syed} M.,  {Blum} J.,  {Wahlberg Jansson} K.,   {Johansen} A.,  2017,
  \mn@doi [\apj] {10.3847/1538-4357/834/2/145}, \href
  {http://adsabs.harvard.edu/abs/2017ApJ...834..145B} {834, 145}

\bibitem[\protect\citeauthoryear{{Chambers}}{{Chambers}}{2014}]{Chambers2014}
{Chambers} J.~E.,  2014, \mn@doi [\icarus] {10.1016/j.icarus.2014.01.036},
  \href {http://adsabs.harvard.edu/abs/2014Icar..233...83C} {233, 83}

\bibitem[\protect\citeauthoryear{{Flynn}, {Consolmagno}, {Brown}  \&
  {Macke}}{{Flynn} et~al.}{2018}]{Flynn2018}
{Flynn} G.~J.,  {Consolmagno} G.~J.,  {Brown} P.,   {Macke} R.~J.,  2018,
  \mn@doi [Chemie der Erde / Geochemistry] {10.1016/j.chemer.2017.04.002},
  \href {https://ui.adsabs.harvard.edu/abs/2018ChEG...78..269F} {78, 269}

\bibitem[\protect\citeauthoryear{{Grott} et~al.,}{{Grott}
  et~al.}{2019}]{Grott2019}
{Grott} M.,  et~al., 2019, \mn@doi [Nature Astronomy]
  {10.1038/s41550-019-0832-x}, \href
  {https://ui.adsabs.harvard.edu/abs/2019NatAs...3..971G} {3, 971}

\bibitem[\protect\citeauthoryear{{Guilera}, {de El{\'{\i}}a}, {Brunini}  \&
  {Santamar{\'{\i}}a}}{{Guilera} et~al.}{2014}]{Guilera2014}
{Guilera} O.~M.,  {de El{\'{\i}}a} G.~C.,  {Brunini} A.,   {Santamar{\'{\i}}a}
  P.~J.,  2014, \mn@doi [\aap] {10.1051/0004-6361/201322061}, \href
  {http://adsabs.harvard.edu/abs/2014A\%26A...565A..96G} {565, A96}

\bibitem[\protect\citeauthoryear{{Gundlach} et~al.,}{{Gundlach}
  et~al.}{2018}]{Gundlach2018}
{Gundlach} B.,  et~al., 2018, \mn@doi [\mnras] {10.1093/mnras/sty1550}, \href
  {http://adsabs.harvard.edu/abs/2018MNRAS.479.1273G} {479, 1273}

\bibitem[\protect\citeauthoryear{{Hornung} et~al.,}{{Hornung}
  et~al.}{2016}]{Hornung2016}
{Hornung} K.,  et~al., 2016, \mn@doi [\planss] {10.1016/j.pss.2016.07.003},
  \href {https://ui.adsabs.harvard.edu/abs/2016P&SS..133...63H} {133, 63}

\bibitem[\protect\citeauthoryear{{Jutzi}, {Michel}, {Benz}  \&
  {Richardson}}{{Jutzi} et~al.}{2010}]{Jutzi2010}
{Jutzi} M.,  {Michel} P.,  {Benz} W.,   {Richardson} D.~C.,  2010, \mn@doi
  [\icarus] {10.1016/j.icarus.2009.11.016}, \href
  {http://adsabs.harvard.edu/abs/2010Icar..207...54J} {207, 54}

\bibitem[\protect\citeauthoryear{{Jutzi}, {Holsapple}, {W{\"u}nneman}  \&
  {Michel}}{{Jutzi} et~al.}{2015}]{Jutzi2015}
{Jutzi} M.,  {Holsapple} K.,  {W{\"u}nneman} K.,   {Michel} P.,  2015,
  preprint, \href {http://adsabs.harvard.edu/abs/2015arXiv150201844J} {}
  (\mn@eprint {arXiv} {1502.01844})

\bibitem[\protect\citeauthoryear{{Kimura} et~al.,}{{Kimura}
  et~al.}{2020}]{Kimura2020}
{Kimura} H.,  et~al., 2020, \mn@doi [\mnras] {10.1093/mnras/staa1641}, \href
  {https://ui.adsabs.harvard.edu/abs/2020MNRAS.496.1667K} {496, 1667}

\bibitem[\protect\citeauthoryear{{Kothe}, {Blum}, {Weidling}  \&
  {G{\"u}ttler}}{{Kothe} et~al.}{2013}]{Kothe2013}
{Kothe} S.,  {Blum} J.,  {Weidling} R.,   {G{\"u}ttler} C.,  2013, \mn@doi
  [\icarus] {10.1016/j.icarus.2013.02.034}, \href
  {https://ui.adsabs.harvard.edu/abs/2013Icar..225...75K} {225, 75}

\bibitem[\protect\citeauthoryear{{Krivov}, {Ide}, {L{\"o}hne}, {Johansen}  \&
  {Blum}}{{Krivov} et~al.}{2018}]{Krivov2018}
{Krivov} A.~V.,  {Ide} A.,  {L{\"o}hne} T.,  {Johansen} A.,   {Blum} J.,  2018,
  \mn@doi [\mnras] {10.1093/mnras/stx2932}, \href
  {https://ui.adsabs.harvard.edu/abs/2018MNRAS.474.2564K} {474, 2564}

\bibitem[\protect\citeauthoryear{Li \& Wong}{Li \& Wong}{2013}]{LiWong2013}
Li D.,  Wong L. N.~Y.,  2013, \mn@doi [Rock Mechanics and Rock Engineering]
  {10.1007/s00603-012-0257-7}, 46, 269

\bibitem[\protect\citeauthoryear{{Macke}, {Consolmagno}  \& {Britt}}{{Macke}
  et~al.}{2011}]{Macke2011}
{Macke} R.~J.,  {Consolmagno} G.~J.,   {Britt} D.~T.,  2011, \mn@doi
  [Meteoritics and Planetary Science] {10.1111/j.1945-5100.2011.01298.x}, \href
  {http://adsabs.harvard.edu/abs/2011M\%26PS...46.1842M} {46, 1842}

\bibitem[\protect\citeauthoryear{Markides \& Kourkoulis}{Markides \&
  Kourkoulis}{2016}]{M&K2016}
Markides C.,  Kourkoulis S.,  2016, \mn@doi [Journal of Rock Mechanics and
  Geotechnical Engineering] {https://doi.org/10.1016/j.jrmge.2015.09.008}, 8,
  127

\bibitem[\protect\citeauthoryear{{Meisner}, {Wurm}  \& {Teiser}}{{Meisner}
  et~al.}{2012}]{Meisner2012}
{Meisner} T.,  {Wurm} G.,   {Teiser} J.,  2012, \mn@doi [\aap]
  {10.1051/0004-6361/201219099}, \href
  {http://adsabs.harvard.edu/abs/2012A\%26A...544A.138M} {544, A138}

\bibitem[\protect\citeauthoryear{{Morbidelli}, {Bottke}, {Nesvorn{\'y}}  \&
  {Levison}}{{Morbidelli} et~al.}{2009}]{Morbidelli2009}
{Morbidelli} A.,  {Bottke} W.~F.,  {Nesvorn{\'y}} D.,   {Levison} H.~F.,  2009,
  \mn@doi [\icarus] {10.1016/j.icarus.2009.07.011}, \href
  {http://adsabs.harvard.edu/abs/2009Icar..204..558M} {204, 558}

\bibitem[\protect\citeauthoryear{{Ostrowski} \& {Bryson}}{{Ostrowski} \&
  {Bryson}}{2019}]{Ostrowski2019}
{Ostrowski} D.,  {Bryson} K.,  2019, \mn@doi [\planss]
  {10.1016/j.pss.2018.11.003}, \href
  {https://ui.adsabs.harvard.edu/abs/2019P&SS..165..148O} {165, 148}

\bibitem[\protect\citeauthoryear{{Parisi}}{{Parisi}}{2013}]{Parisi2013}
{Parisi} M.~G.,  2013, \mn@doi [\planss] {10.1016/j.pss.2012.09.013}, \href
  {https://ui.adsabs.harvard.edu/abs/2013P&SS...75...96P} {75, 96}

\bibitem[\protect\citeauthoryear{{San Sebasti{\'a}n} \& {Parisi}}{{San
  Sebasti{\'a}n} \& {Parisi}}{2016}]{Sanse2016}
{San Sebasti{\'a}n} I.~L.,  {Parisi} M.~G.,  2016, Boletin de la Asociacion
  Argentina de Astronomia La Plata Argentina, \href
  {http://adsabs.harvard.edu/abs/2016BAAA...58..307S} {58, 307}

\bibitem[\protect\citeauthoryear{{San Sebasti{\'a}n}, {Guilera}  \&
  {Parisi}}{{San Sebasti{\'a}n} et~al.}{2019}]{SanSebastian2019}
{San Sebasti{\'a}n} I.~L.,  {Guilera} O.~M.,   {Parisi} M.~G.,  2019, \mn@doi
  [\aap] {10.1051/0004-6361/201834168}, \href
  {https://ui.adsabs.harvard.edu/abs/2019A&A...625A.138S} {625, A138}

\bibitem[\protect\citeauthoryear{{Slyuta}}{{Slyuta}}{2017}]{Slyuta2017}
{Slyuta} E.~N.,  2017, \mn@doi [Solar System Research]
  {10.1134/S0038094617010051}, \href
  {https://ui.adsabs.harvard.edu/abs/2017SoSyR..51...64S} {51, 64}

\bibitem[\protect\citeauthoryear{{Weidenschilling}}{{Weidenschilling}}{2011}]{Weidenschilling2011}
{Weidenschilling} S.~J.,  2011, \mn@doi [\icarus]
  {10.1016/j.icarus.2011.05.024}, \href
  {http://adsabs.harvard.edu/abs/2011Icar..214..671W} {214, 671}

\makeatother
\end{thebibliography}
\bibliographystyle{mnras}




\appendix




\bsp	
\label{lastpage}
\end{document}